\documentclass[%
aps,reprint,superscriptaddress,prl,
amsmath,amssymb,
nofootinbib,nobibnotes,
]{revtex4-2}
\usepackage{titlesec}
\usepackage{graphicx}
\usepackage{dcolumn}
\usepackage{bm}
\usepackage{hyperref}
\hypersetup{
    colorlinks,
    linkcolor={blue},
    citecolor={blue},
    urlcolor={blue}
}

\usepackage[separate-uncertainty=true,range-phrase=--,multi-part-units=single,range-units=single]{siunitx}
\usepackage{orcidlink}

\newcommand{\SM}[2]{\hyperref[#1]{#2} in Supplemental Materials}
\newcommand{\avg}[1]{\langle#1\rangle}
\DeclareMathOperator{\sign}{sign}
\DeclareMathOperator{\const}{const}
\allowdisplaybreaks[1]

\begin{document}


\title{Poissonian cellular Potts models reveal nonequilibrium kinetics of cell sorting}

\author{R. Belousov \orcidlink{0000-0002-8896-8109}}
\email{roman.belousov@embl.de}
\affiliation{Cell Biology and Biophysics Unit, European Molecular Biology Laboratory, Meyerhofstraße 1, 69117 Heidelberg, Germany}
\author{S. Savino \orcidlink{0009-0004-6938-2219}}
\affiliation{Cell Biology and Biophysics Unit, European Molecular Biology Laboratory, Meyerhofstraße 1, 69117 Heidelberg, Germany}
\affiliation{Department of Mathematical Sciences, Politecnico di Torino, Corso Duca degli Abruzzi 24, 10129 Turin, Italy}
\author{P. Moghe \orcidlink{0000-0002-7924-9671}}
\affiliation{Hubrecht Institute, Uppsalalaan 8, 3584 CT Utrecht, Netherlands}
\affiliation{Developmental Biology Unit, European Molecular Biology Laboratory, Meyerhofstraße 1, 69117 Heidelberg, Germany. Collaboration for joint PhD degree between EMBL and Heidelberg University, Faculty of Biosciences, Heidelberg, Germany.}
\author{T. Hiiragi \orcidlink{0000-0003-4964-7203}}
\affiliation{Hubrecht Institute, Uppsalalaan 8, 3584 CT Utrecht, Netherlands}
\affiliation{Institute for the Advanced Study of Human Biology (WPI-ASHBi), Kyoto University, Kyoto, Japan}
\affiliation{Department of Developmental Biology, Graduate School of Medicine, Kyoto University, Kyoto, 606-8501, Japan}
\author{L. Rondoni \orcidlink{0000-0002-4223-6279}}
\affiliation{Department of Mathematical Sciences, Politecnico di Torino, Corso Duca degli Abruzzi 24, 10129 Turin, Italy}
\affiliation{INFN, Sezione di Torino, Turin 10125, Italy}
\author{A. Erzberger \orcidlink{0000-0002-2200-4665}}%
\email{erzberge@embl.de}
\affiliation{Cell Biology and Biophysics Unit, European Molecular Biology Laboratory, Meyerhofstraße 1, 69117 Heidelberg, Germany}
\affiliation{Department of Physics and Astronomy, Heidelberg University, 69120 Heidelberg, Germany}




\date{\today}

\begin{abstract}
Cellular Potts models are broadly applied across developmental biology and cancer research. We overcome limitations of the traditional approach, which reinterprets a modified Metropolis sampling as ad hoc dynamics, by introducing a physical timescale through Poissonian kinetics and by applying principles of stochastic thermodynamics to separate thermal and relaxation effects from athermal noise and nonconservative forces. Our method accurately describes cell-sorting dynamics in mouse-embryo development and identifies the distinct contributions of nonequilibrium processes, e.g. cell growth and active fluctuations.
\end{abstract}

\maketitle

The dynamics of many nonequilibrium systems can be described by a time-dependent phenomenological Hamiltonian which actively controls transitions through a sequence of target states. Widely adopted frameworks of vertex, cellular Potts, and other methods rely on this effective energy-based principle to explain spatial organization in living systems~\cite{Tanaka2015,Fletcher2017,Alt2017,granerSimulationBiologicalCell1992,Glazier1993,JNewman2005,Christley2010,Fletcher2013,Okuda2015,Revell2019,durandLargescaleSimulationsBiological2021}.

Whereas the optimum of the system's energy specifies a target state of such an active transformation, the unfolding of the modeled process in time is determined by its kinetic parameters. In vertex or subcellular-element models with a continuous phase space, these parameters correspond to the transport properties---the damping coefficients. However the traditional cellular Potts models (CPMs), whose discrete-state dynamics are implemented by a modified Metropolis sampling, lack an explicit control over such kinetic parameters.

Transport properties and the timescales they control are especially important when multiple processes evolve interdependently. In the course of embryonic development, numerous cellular and tissue-level processes require precise mutual coordination~\cite{Negrete2021}. For example the sorting of cell types in the early mouse embryo must be completed before the subsequent morphogenetic events commence \cite{Kojima2014,Bondarenko2022}.

To introduce kinetic parameters into CPMs we invoke the theory of stochastic thermodynamics~\cite{vandenbroeckc.StochasticThermodynamicsBrief2013,pelitiStochasticThermodynamicsIntroduction2021}, which comprehensively describes discrete-state physical processes driven by changes of free energy. As shown further, the transport properties control the system's \textit{frenetic activity}~\cite{Maes2020}, which constitutes the time-symmetric component of a \textit{stochastic action}---the complement of entropic changes of a system's trajectory.

\begin{figure}[!t]
\includegraphics[width=\columnwidth]{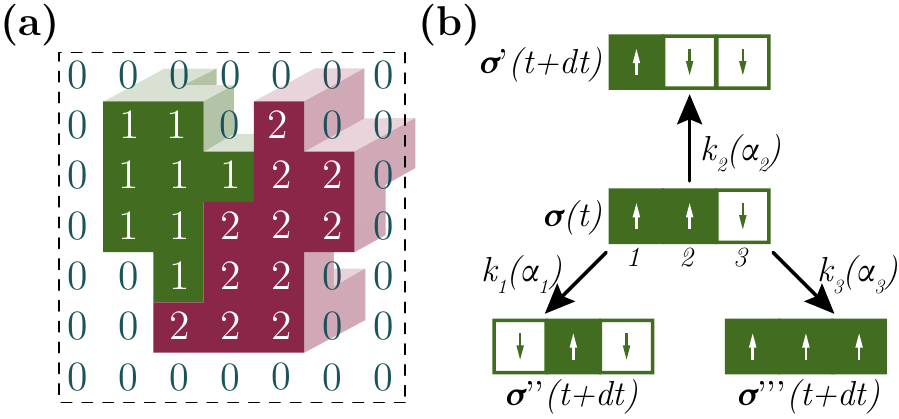}
\caption{\label{fig:sketch}(a)~Schematic of a CPM: Simply connected regions of a voxel grid with ``spin'' values 1 (green) and 2 (red) represent two cells in a medium with value 0. (b)~Poissonian dynamics of three Ising spins $\bm{\sigma}=(\sigma_1, \sigma_2, \sigma_3)$: The system's configuration $\bm{\sigma}(t)$ may change within a small time $dt$ into one of the target states ($\bm{\sigma}'$, $\bm{\sigma}''$, $\bm{\sigma}'''$), which differ from the original one by the value of a single spin $\sigma_{i=1,2,3}$, because Poissonian events never occur simultaneously.
}
\end{figure}

While being less demanding than the subcellular-element method, CPMs can treat composite materials and describe more intricate shapes than vertex models~\cite{granerSimulationBiologicalCell1992,savillModellingMorphogenesisSingle1997,sciannaMultiscaleDevelopmentsCellular2012,voss-bohmeMultiScaleModelingMorphogenesis2012,hirashimaCellularPottsModeling2017}. Each cell in three dimensions corresponds to a contiguous collection of voxels with the same ``spin'' value---labels distinguishing individual objects in the system [Fig.~\ref{fig:sketch}(a)].

\begin{figure*}[!t]
\includegraphics[width=0.9\textwidth]{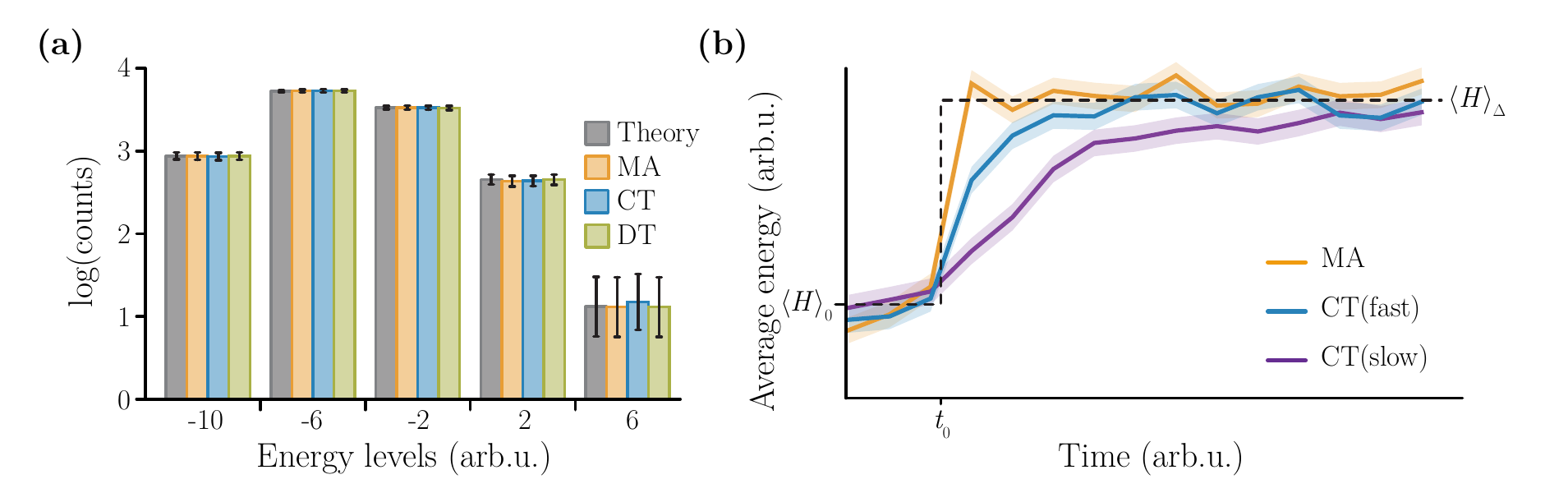}
\caption{\label{fig:ising}Simulations of an Ising chain with $N = 10$ spins. (a) Distribution of the total energy in the canonical ensemble at $T = 2 J / k_{\rm{B}}$ ($J = 1$ arb.u.). Tau-leap simulations in discrete time (DT, $dt=10^{-4}$ arb.u.) match the exact distribution (Theory); p-value of the multinomial test $0.997$~\cite{lawleyGENERALMETHODAPPROXIMATING1956,williamsImprovedLikelihoodRatio1976}. The continuous-time simulations (CT) and the Metropolis algorithm (MA) produce comparable results (p-values $0.966$ and $0.986$ respectively). In the DT and CT models the chain is inhomogeneous with action rates ($\alpha_{2 i + 1}=0.1$ for the odd indices, $\alpha_{2 i}=0.3$ arb.u. for the even). Error bars are given by three standard deviations of $10^4$ realizations. (b) Relaxation of the chain energy between two equilibrium states with average values $\avg{H}_0$ and $\avg{H}_\Delta$ at temperatures $T(t < t_0) = T_0 = 1.8 |J|$ and $T(t \gg t_0) = T_0 + \Delta{T} = 2.0 |J| / k_{\rm{B}}$ respectively ($J = -1$ arb.u.). The results of MA sampling are reported alongside the CT simulations of a slow ($\alpha_{2 i} = 0.05$ and $\alpha_{2 i + 1} =0.08$ arb.u.) and fast ($\alpha_{2 i} = 0.5$ and $\alpha_{2 i + 1} =0.7$ arb.u.) kinetics. Each curve traces an average over $10^4$ trajectories with a standard-error band.}
\end{figure*}

CPMs were first introduced by \citet{granerSimulationBiologicalCell1992} to study how differences of surface energy between homotypic and heterotypic contacts cause cell sorting in development. Using the modified Metropolis algorithm they found that clusters of cells emerge in a typical configuration favored by the system's energy function
\begin{equation}\label{eq:E}
    E = \sum_{ij} \frac{J_{ij}(\sigma_i, \sigma_j)}{2} + \sum_k \frac{\kappa_k (V_k - \bar{V}_k)^2}{2},
\end{equation}
in which the first sum runs over spin pairs $\sigma_i$ and $\sigma_j$ with symmetric coefficients $J_{ij}(\sigma_i,\sigma_j) = J_{ji}(\sigma_i,\sigma_j)$ encoding the surface interactions, whereas the second term penalizes deviations of the volume $V_k$ of the $k$\textsuperscript{th} cell from its preferred value $\bar{V}_k$. Usually $J_{ij}(\sigma_i,\sigma_j)$ are identically zero unless the spins $\sigma_i$ and $\sigma_j$ are in direct contact. As the method of \citet{granerSimulationBiologicalCell1992} evolved beyond a mere proof of concept, it was further generalized to include nonequilibrium aspects, such as cell division and active motility \cite{granerSimulationBiologicalCell1992,savillModellingMorphogenesisSingle1997,sciannaMultiscaleDevelopmentsCellular2012,voss-bohmeMultiScaleModelingMorphogenesis2012,hirashimaCellularPottsModeling2017,chenParallelImplementationCellular2007,andasariIntegratingIntracellularDynamics2012,sciannaCellularPottsModel2013,szaboCellularPottsModeling2013,cerrutiPolarityCellDivision2013,osborneMultiscaleModelColorectal2015,belmonteVirtualtissueComputerSimulations2016,durandEfficientCellularPotts2016,thuroffBridgingGapSinglecell2019,berghoffCellsSilicoIntroducing2020,durandLargescaleSimulationsBiological2021,Kabla2012,Guisoni2018,Beatrici2022,nakajimaKineticsCellularPotts2011,voss-bohmeMultiScaleModelingMorphogenesis2012,Mare,Magno2015,rensEnergyCellularForces2019}.

\begin{figure*}[!t]
\includegraphics[width=\textwidth]{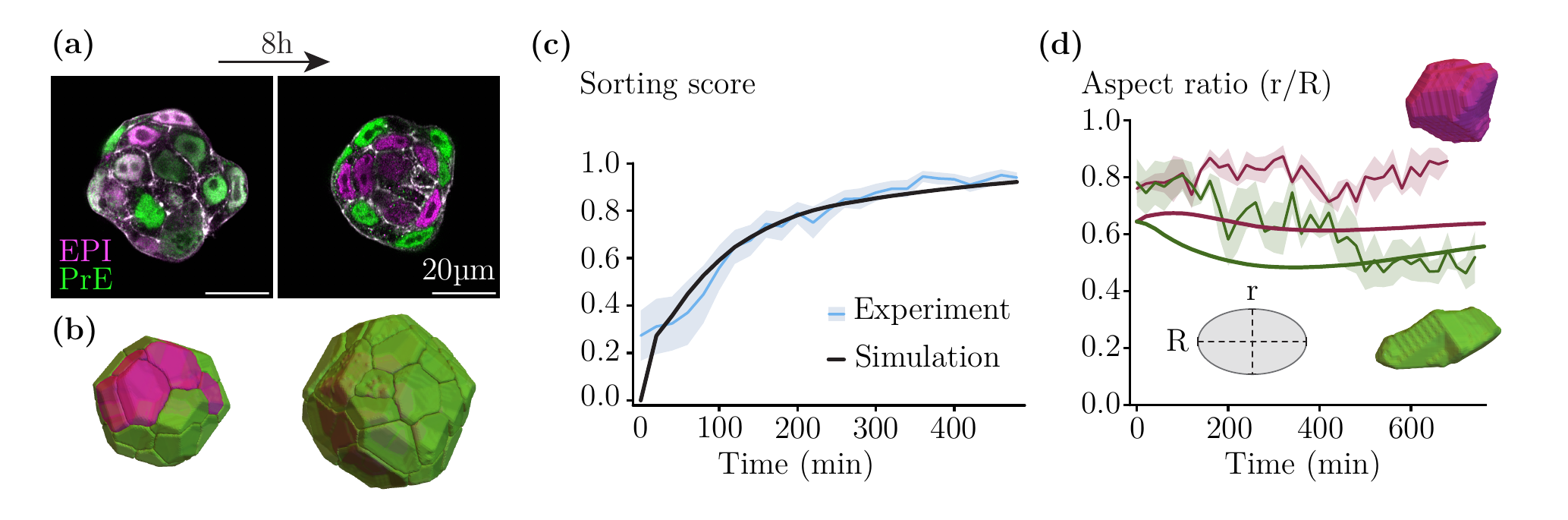}
\caption{\label{fig:data}Sorting of epiblast (EPI) and primitive endoderm (PrE) cells in ICMs isolated from  mouse blastocysts. (a) Immunostaining images on embryonic day 3.5 and \SI{8}{h} later (see \SM{sec:exp}{Sec. 4}). (b)~Cellular-Potts simulation in the initial configuration and \SI{8}{h} later. (c)~Sorting score of live-imaging data averaged over 8 experiments compared to a mean trend of 500 simulations [$\alpha_{0, {\rm EPI}, {\rm PRE}} = (0.95, 0.69, 2.12)\,\si{min^{-1}}$]; standard-error band shown only for the experiments. (d) Average cellular aspect ratios characterized in the maximum-projection planes of experimental data and by the eigenvalues of the 3D gyration tensors in the simulations.}
\end{figure*}

The modified Metropolis algorithm of modern CPMs is not the most general kinetic model for discrete systems~\cite{Glauber1963,Bortz1975,Prados1997,Serebrinsky2011,Zeegers2013} and has limitations~\cite{Glazier,durandEfficientCellularPotts2016}. In fact, the Metropolis scheme was originally designed to bypass slow simulations of systems' dynamics when sampling equilibrium ensembles \cite[Chapter 7]{Metropolis1953,tuckermanStatisticalMechanicsTheory2010}. Since the early days of CPMs, questions have therefore been raised regarding the interpretation of time, temperature, dissipation effects, and nonequilibrium aspects of this approach~\cite{Glazier}.

As discrete-state Markovian systems, which describe continuous-time physical phenomena, CPMs can be most naturally regarded as Poissonian processes (see \SM{sec:cpm}{Sec. 1}). The framework proposed here leverages this result of queuing theory \cite{Nelsen1987} to address the above questions. Through Poissonian kinetics and stochastic thermodynamics we introduce interpretable time and energy scales, account for response coefficients and forces not incorporated in the Hamiltonian, and finally separate thermal and athermal fluctuations.

\textit{Framework}.---As an illustration of our approach we first consider a paradigmatic example of discrete systems and a special case of the Potts model~\cite{wuPottsModel1982}---an Ising chain $\bm{\sigma}=(\sigma_1, \sigma_2,...,\sigma_N)$ with nearest-neighbor interactions. Arranged on a one-dimensional lattice, $N$ spins $\sigma_{i=1,2...N} \in \{-1,1\}$ with a periodic boundary condition $\sigma_{N+1} = \sigma_1$ are described by the Hamiltonian
$$
    H = \sum_{i=1}^N \frac{J}{2} \sigma_i \sigma_{i+1}
$$
with an interaction constant $J$.

To define dynamics of the Ising chain we assume that each spin flips its sign with a Poissonian transition rate $k_i(\bm{\sigma})$, which in general depends on the current state $\bm{\sigma}$~[Fig.~\ref{fig:sketch}(b)]. Within a sufficiently small time $dt$ at most one spin can change its value. In equilibrium, the detailed-balance condition for such a spin $\sigma_i$ requires
\begin{equation}\label{eq:balance}
    \exp\left[
        -\frac{H(\sigma_i)}{k_{\rm{B}} T}
    \right] k_i(\sigma_i)
    = \exp\left[
        -\frac{H(-\sigma_i)}{k_{\rm{B}} T}
    \right] k_i(-\sigma_i),
\end{equation}
in which $H(\sigma_i)$ and $k_i(\sigma_i)$ are respectively the spin's energy and transition rate, given the values of $\sigma_{j\ne i}$. From Eq.~\eqref{eq:balance} it follows then
\begin{equation}\label{eq:ratio}
  \frac{k_i(\sigma_i)}{k_i(-\sigma_i)} = \exp\left[-\frac{\Delta{H}(-\sigma_i)}{k_{\rm{B}} T}\right]
\end{equation}
with $\Delta{H}(-\sigma_i) = H(-\sigma_i) - H(\sigma_i)$.

The chain's stochastic kinetics is then given by a master equation, once a common factor between $k_i(\sigma_i)$ and $k_i(-\sigma_i)$ in Eq.~\eqref{eq:ratio} is specified~\cite{Glauber1963,herpichCollectivePowerMinimal2018,freitasStochasticThermodynamicsNonlinear2021,meibohmFiniteTimeDynamicalPhase2022}. In a general context each spin may be characterized by a state-dependent \textit{action rate} $\alpha_i(\sigma_i)$, which determines the probability
$$1 - e^{-\alpha_i(\sigma_i) dt} \approx \alpha_i(\sigma_i) dt$$
for the $i$\textsuperscript{th} spin to \textit{attempt} a sign change. When such an attempt occurs, the transition probability $p(\sigma_i\to\sigma_i')$ is determined by a \textit{directing function} $L(\sigma_i')$ with a normalization constant $Z$~\cite{Belousov2022}
\begin{equation}\label{eq:transition}
    p(\sigma_i\to\sigma_i') = \frac{e^{L(\sigma_i')}}{Z}.
\end{equation}
For the two possible outcomes---no change, $\sigma_i'=\sigma_i$, and a transition, $\sigma_i'=-\sigma_i$---the normalization constant expands to $Z = e^{L(\sigma_i)} + e^{L(-\sigma_i)}$ and Eq.~\eqref{eq:transition} yields
\begin{align}
    p(\sigma_i\to\sigma_i) =& \frac{1}{1 + e^{\Delta{L}(-\sigma_i)}},
    \\\label{eq:success}
    p(\sigma_i\to-\sigma_i) =& \frac{e^{\Delta{L}(-\sigma_i)}}{1 + e^{\Delta{L}(-\sigma_i)}},
\end{align}
with $\Delta{L}(-\sigma_i) = L(-\sigma_i) - L(\sigma_i)$. With the transition rate given by the product of the attempt rate and the transition probability \begin{equation}\label{eq:rate}
    k_i(\sigma_i) = \alpha_i(\sigma_i) \frac{e^{\Delta{L}(-\sigma_i)}}{1 + e^{\Delta{L}(-\sigma_i)}},
\end{equation} we find from Eq.~\eqref{eq:ratio}
\begin{equation}\label{eq:weight}
    e^{\Delta{L}(-\sigma_i)} = \frac{\alpha_i(-\sigma_i)}{\alpha_i(\sigma_i)} \exp\left[-\frac{\Delta{H}(-\sigma_i)}{k_{\rm{B}} T}\right].
\end{equation}

A complete specification of the Ising chain now requires both the Hamiltonian and the spins' action rates. Such a system can be simulated exactly in continuous time by the standard techniques for master equations, or approximately by using a tau-leap algorithm with a step $dt$~\cite{Gillespie2001}.

The action rates do not compromise the canonical distribution of the Ising chain in equilibrium [Fig.~\ref{fig:ising}(a)]. These parameters control the unfolding of dynamical processes, such as relaxation of transients. For example, chains with larger action rates, initially prepared in equilibrium at temperature $T_0$ and subject to a sudden temperature change $\Delta{T}$, relax to the new steady state faster~[Fig.~\ref{fig:ising}(b)]. By design, the Metropolis scheme renders samples of the target equilibrium ensemble after a very short transient trajectory, which can not be controlled by algorithmic or system parameters.

\textit{Model analysis}.---To analyze the Poissonian dynamics of the Ising chain we apply the theory of stochastic thermodynamics. Any given trajectory of the system $\theta$ from an initial state $\bm{\sigma}^0$ to a final state $\bm{\sigma}^M$ can be decomposed into a sequence of $M$ \textit{elementary paths}
$$\theta = \mathcal{T}_M \mathcal{T}_{M-1} ... \mathcal{T}_1.$$
Each path $\mathcal{T}$ consists of $n$ quiescent intervals of arbitrarily small time $dt$ in the same configuration $\bm{\sigma}$, followed by a sign change of the $i$\textsuperscript{th} spin which produces the next state $\bm{\sigma}'$. The probability of this change is $p_i \simeq k_i(\sigma_i) dt$, whereas the probability of the quiescent period lasting $n$ steps is
\begin{equation}
    q_n = \left[1 - dt \sum_j k_j(\bm{\sigma})  \right]^n \approx e^{-n dt \sum_j k_j(\bm{\sigma})}.
\end{equation}
With these definitions, the probability of the elementary path from a given initial condition is
\begin{equation}
    p(\mathcal{T} | \bm{\sigma}) = q_n p_i. 
\end{equation}

Now we can decompose the \textit{stochastic action} $\mathcal{A}$ of the elementary path into the entropic and frenetic components~\cite{Maes2020}, $\Delta\mathcal{S}$ and $\mathcal{D}$ respectively:
\begin{equation}
    \mathcal{A} = -k_{\rm{B}} \ln p(\mathcal{T} | \bm{\sigma}) = \mathcal{D} - \frac{1}{2}\Delta\mathcal{S}.
\end{equation}
Indeed the probability of a time-reverse trajectory $\tilde{\mathcal{T}}$---a change of the $i$\textsuperscript{th} spin conditioned on the initial configuration $\bm{\sigma}'$ and followed by $n$ quiescent steps---is
\begin{equation}
    p(\tilde{\mathcal{T}} | \bm{\sigma}') = p_i' q_n
\end{equation}
in which $p_i' = k_i(-\sigma_i) dt$. Due to Eqs.~\eqref{eq:rate} and \eqref{eq:weight} the time-asymmetric part of the action yields
\begin{align}\label{eq:S}
    \Delta\mathcal{S} =& -k_{\rm{B}} \ln \frac{p(\mathcal{T} | \bm{\sigma})}{p(\tilde{\mathcal{T}} | \bm{\sigma}')}
        \nonumber\\
        =& -k_{\rm{B}} \ln \frac{k_i(\sigma_i)}{k_i(-\sigma_i)} = \frac{\Delta{H}(-\sigma_i)}{T}.
\end{align}
The time-symmetric part renders a more involved expression approximated by
\begin{align}\label{eq:D}
    \mathcal{D} =& \frac{k_{\rm{B}}}{2} \ln \left[ p(\mathcal{T} | \bm{\sigma})p(\tilde{\mathcal{T}} | \bm{\sigma}') \right]
        \approx - n dt k_{\rm{B}} \sum_j k_j(\sigma_i) \nonumber\\&+ k_{\rm{B}} \ln [\sqrt{k_i(\sigma_i) k_i(-\sigma_i)} dt]
\end{align}
for small $dt$. The total action of the whole trajectory is  $\mathcal{A}(\theta) = \sum_{m=0}^M \mathcal{A}(\mathcal{T}_m)$ with the components
\begin{align}\label{eq:entropy}
    \Delta\mathcal{S}(\theta) =& \frac{1}{T} \left\{
            H\left[\bm{\sigma}^{M}\right] - H(\bm{\sigma}^{0})
    \right\},
    \\\label{eq:frenesy}
    \mathcal{D}(\theta) =& \sum_{m=1}^M \mathcal{D}(\mathcal{T}_m).
\end{align}

Without compromising the entropic activity, action rates control the system's frenesy through transition rates $k_j$, cf. Eqs~\eqref{eq:rate},  \eqref{eq:D}, and \eqref{eq:frenesy}. The entropy change of a relaxation process is entirely determined by the energy difference between the initial and final configurations of the system [Eq.~\eqref{eq:S}, Fig.~\ref{fig:ising}(b)]. In contrast, the frenesy depends on the kinetics of each state transition in a system's trajectory.


\begin{figure*}[t]
\includegraphics[width=\textwidth]{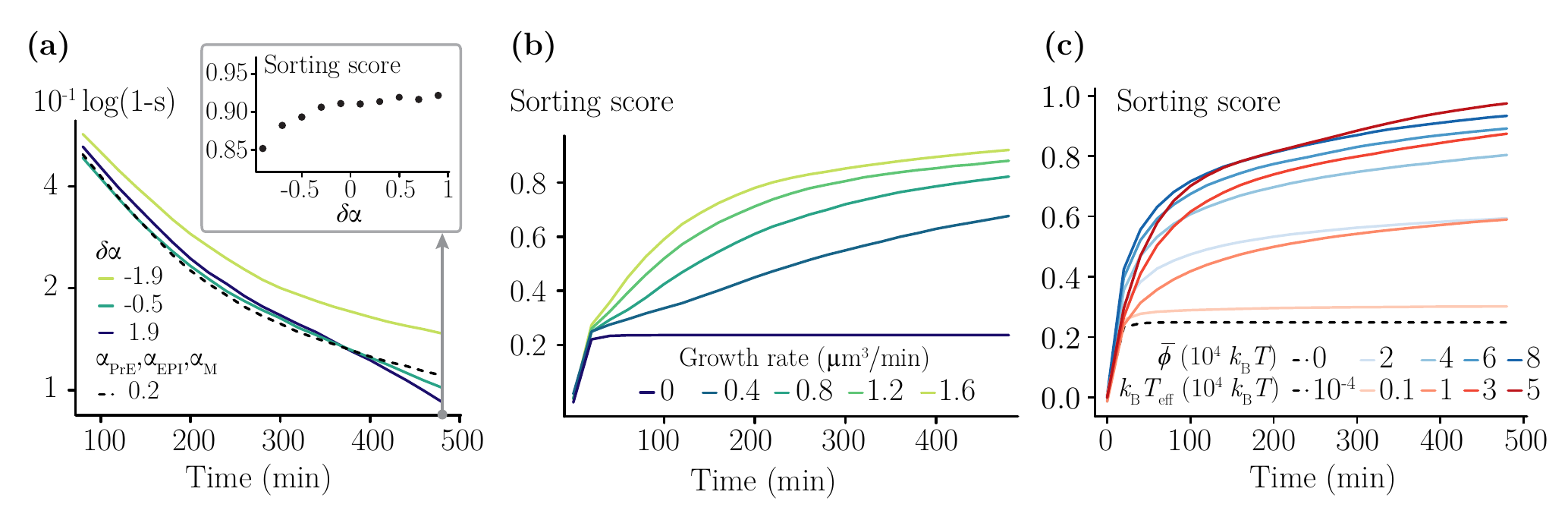}
\caption{\label{fig:pars}Kinetic and nonequilibrium aspects of Poissonian CPMs. (a) Convergence of sorting $\log(1-s)$ for different values of heterogeneity $\delta\alpha = (\alpha_{\rm PrE} - \alpha_{\rm Epi}) / \alpha_0$, $\alpha_0=\SI{1}{min^{-1}}$]. Inset: Scores at $t=\SI{480}{min}$ show that faster PrE kinetics promotes sorting. Heterogeneous kinetic properties produce non-rescalable differences in the sorting dynamics---note the two-point intersection between the black dashed curve with homogeneous action rates ($\alpha_0 = \alpha_{\rm EPI} = \alpha_{\rm PrE} = \SI{0.2}{min^{-1}}$) and the green curve ($\delta{\alpha} = \SI{-0.5}{min^{-1}}$). (b--c) Either cell growth and division, or active fluctuations ($\bar{\phi}$), or extreme values of an \textit{effective temperature} $k_B T_{\rm e}$ are required for complete sorting. Curves in panels (a) and (b--c) represent averages over 1000 and 500 trajectories respectively [parameters set to values as in Fig.~3(c) unless noted otherwise]. 
}
\end{figure*}

\textit{Poissonian cellular Potts models}.---We now construct a three-dimensional Poissonian CPM~\cite{dycpm}. Voxels on a cubic lattice describe the state of $K$ distinct cells and a medium, taking values $\sigma_i \in \{0,1,2,...,K\}$ (Fig.~\ref{fig:sketch}). The coefficients $J_{ij}(\sigma_i, \sigma_j)$ in Eq.~\eqref{eq:E} vanish when voxels $i$ and $j$ are both occupied by the same object, or when the voxel $i$ is not within the Moore neighborhood of the voxel $j$~\cite{durandEfficientCellularPotts2016}. Otherwise $J_{ij}$ assume constant positive values encoding the surface interactions between objects.

For each of the total $\nu$ object \textit{types}, our framework introduces a Poissonian state-dependent action rate $\alpha(\sigma_i) \in \{\alpha_0, \alpha_\mathrm{1}, ..., \alpha_\nu\}$, with which an $i$\textsuperscript{th} voxel attempts to change its current value $\sigma_i$. Its possible target values $\sigma_i^{(j)}$ are chosen from the Von Neumann neighborhood like in the standard CPMs~\cite{durandEfficientCellularPotts2016}, with the transition probabilities given by a general version of Eqs.~\eqref{eq:transition}~and~\eqref{eq:weight}
\begin{align}\label{eq:cpmsuccess}
    p(\sigma_i \to \sigma_i^{(j)}) =& \frac{e^{\Delta{L}(\sigma_i^{(j)})}}{\sum_j e^{\Delta{L}(\sigma_i^{(j)})}},
    \\\label{eq:cpmweight}
    e^{\Delta{L}(\sigma_i^{(j)})} =& \frac{\alpha(\sigma_i^{(j)})}{\alpha(\sigma_i)} \exp\left[-\frac{H(\sigma_i^{(j)}) - H(\sigma_i)}{k_{\rm{B}} T}\right].
\end{align}

Temperature in traditional CPMs is a fictitious parameter manipulated to adjust the level of fluctuations~\cite{Swat2012}. In contrast, our approach regards it as a physical variable that is set at an experimentally controlled value. To prevent cell fragmentation, usually suppressed by a periodically applied annealing, we adopt the local-connectivity test of \citet{durandEfficientCellularPotts2016} in a modified form (see \SM{sec:cpm}{Sec.~3}).

An extension of the directing-function formalism~\cite[Appendix B]{Belousov2022} can also incorporate more general nonequilibrium forces into Eq.~\eqref{eq:cpmsuccess} as
\begin{equation}\label{eq:active}
 p(\sigma_i \to \sigma_i^{(j)}) = \frac{e^{\Delta{L}(\sigma_i^{(j)}) + \phi(\sigma_i, \sigma_i^{(j)})}}{\sum_j e^{\Delta{L}(\sigma_i^{(j)}) + \phi(\sigma_i, \sigma_i^{(j)})}},
\end{equation}
in which the \textit{active exponents} $\phi(\sigma_i, \sigma_i^{(j)})$ are functions associated with a specific transition $\sigma_i \to \sigma_i^{(j)}$, and in general $\phi(\sigma_i, \sigma_i^{(j)}) \ne \phi(\sigma_i^{(j)}, \sigma_i)$. When this perturbation breaks the detailed-balance condition, such a transition incurs an irreversible thermodynamic work.

Active exponents can introduce non-Hamiltonian forces and, by violating the fluctuation-dissipation theorem, athermal noise (see \SM{sec:phi}{Sec.~2}).
Here we focus on noise amplification by nonequilibrium processes, which are usually present inside cells, such as the chemically-driven polymerization of cytoskeletal filaments or molecular motor activity~\cite{Kruse2005,Julicher2007,Laplaud2021}. If we set
\begin{equation}\label{eq:phi}
    \phi(\sigma_i, \sigma_i) \equiv 0,\qquad
    \phi(\sigma_i, \sigma_i^{(j)} \ne \sigma_i) = \bar{\phi} = \const,
\end{equation}
all transitions $\sigma_i\to\sigma_i^{(j)}$, except for the trivial ones $\sigma_i^{(j)} \equiv \sigma_i$ are promoted. Nonconservative forces are not generated by active exponents, whose asymmetric components vanish $\phi(\sigma_i, \sigma_i^{(j)}) - \phi(\sigma_i^{(j)}, \sigma_i) = 0$.

Specific active processes can be modeled more explicitly as well. Cell growth is typically implemented by a time-dependent preferred volume $\bar{V}_k$ in Eq.~\eqref{eq:E}. Persistent cell motility can be incorporated by additional terms of the Hamiltonian~\cite{Belmonte2008,Kabla2012,Guisoni2018,Beatrici2022,thuroffBridgingGapSinglecell2019}, or by asymmetric active exponents.


\textit{Cell sorting during embryonic development}.---As a biophysical example we consider the sorting of epiblast (EPI) and primitive endoderm (PrE) cells in the early mouse embryo. These cells form the inner cell mass (ICM) aggregate and sort into an outer single layer of PrE cells separating the epiblast from the medium~\cite{Saiz2020} [Fig.~\ref{fig:data}(a)].

Recent advances provide unprecedented experimental access to the dynamics of isolated ICMs~\cite{Solter1975,Wigger2017,Kim2022,Yanagida2022,companion}. We quantify the segregation of the two cell types by a \textit{sorting score} computed from distances of EPI and PrE cells ($r^\mathrm{EPI}_i$ and $r^\mathrm{PrE}_j$ respectively) from their common geometric center:
$$
    s = \frac{1}{N^\mathrm{PrE}\ N^\mathrm{EPI}}\sum_{i=1}^{N^\mathrm{EPI}} \sum_{j=1}^{N^\mathrm{PrE}} \sign\left(r^\mathrm{PrE}_j - r^\mathrm{EPI}_i\right),
$$
in which $N^\mathrm{PrE}$ and $N^\mathrm{EPI}$ are the numbers of PrE and EPI cells. By definition the score $s \in [-1, 1]$ is close to zero for unsorted cells, and $-1$ or $1$ when all PrE cells are inside or outside the aggregate respectively. To model the sorting process, we chose the five interaction constants $J_{\mathrm{medium}:\mathrm{EPI}}$, $J_\mathrm{\mathrm{medium}:PrE}$, $J_{\mathrm{EPI}:\mathrm{EPI}}$, $J_{\mathrm{EPI}:\mathrm{PrE}}$, $J_{\mathrm{PrE}:\mathrm{PrE}}$ from a physiologically relevant range of the EPI and PrE surface tensions, set the temperature to the experimental value at \SI{310.15}{K}, and calibrated the growth parameters to match the observed proliferation dynamics (see \SM{sec:num}{Sec.~3}).

Almost perfect sorting is achieved within \SI{480}{min} of CPM simulations for a wide range of parameters [Fig.~\ref{fig:data}(b)]. The action rates control the relaxation dynamics of the sorting process [Fig.~\ref{fig:data}(c), \SM{fig:si}{Fig.~S1(b)}]. We sampled 100 combinations of the values $\{\alpha_0, \alpha_\mathrm{EPI}, \alpha_\mathrm{PrE}\}$, with each entry chosen from the interval $(0.10, 3.57)$ $\si{min^{-1}}$. The best match is closest to the experimental curve in the least-squares sense.

Furthermore, our simulations predict distinct shape dynamics of the two cell types, which quantitatively agree  with the experimental data in a parameter-free comparison of cellular aspect ratios [Fig.~\ref{fig:data}(d)]: EPI cells tend to more rounded shapes, whereas PrE cells stretch normally to the radial direction at the outermost shell of ICM. For more details on ICM sorting \emph{in vivo} see Ref.~\cite{companion}.

Faster kinetics of PrE cells promote sorting [Fig.~\ref{fig:pars}(a)]. This effect of inhomogeneous action rates can neither be modeled in the traditional CPMs (see \SM{sec:cpm}{Sec.~1}), nor can it be compensated by time rescaling, as the curves parameterized by $\alpha_{0,{\rm EPI},{\rm PrE}}$ in general belong to different families [Fig.~\ref{fig:pars}(a)].

Without cell growth and division [Fig.~\ref{fig:pars}(b)], or active fluctuations [Fig.~\ref{fig:pars}(c)] sorting is hindered. Both mechanisms do a thermodynamic work on the system: the growth of cells generates stresses, and new cell boundaries increase the total surface energy, whereas the active fluctuations inject energy by breaking detailed balance. Responding to these nonequilibrium processes, the system rapidly acquires the energetically favored sorted state. Modeling active fluctuations by an effective temperature is also viable~\cite{Cugliandolo1997,Polettini2019}, but may misrepresent the system's response to thermodynamic forces~(see \SM{sec:phi}{Sec. 2}).

In fact, the response coefficients of cells' surfaces are directly related to action rates (see \SM{sec:phi}{Sec.~2}). Because energy is a derived unit of time, ``independent'' rescaling of time or temperature, afforded by the traditional CPMs at the account of fictitious energy scales, is forbidden once the experimental data fix $k_B T$ and $J_{ij}$.

\textit{Conclusions}.---Poissonian CPMs provide a physically consistent framework to study complex materials with active properties, which is generally applicable to other discrete-state systems~\cite{Solon2013,herpichCollectivePowerMinimal2018,freitasStochasticThermodynamicsNonlinear2021,meibohmFiniteTimeDynamicalPhase2022}. Its kinetic parameters control transport coefficients and permit an unambiguous interpretation of time. Active fluctuations and nonequilibrium processes are clearly separated from thermal effects and passive relaxation. We applied this framework to examine the roles of distinct nonequilibrium processes in embryonic cell sorting, and show that either growth and division, or active shape fluctuations are required for successful segregation of cell types.

\begin{acknowledgments}
\textit{Acknowledgements}.---R.B. is grateful to Marc Durand for stimulating discussions on the fragmentation-free CPM approach for 3D systems and to Florian Berger for creative suggestions on the front matter. R.B, S.S, P.M, T.H., and A.E. acknowledge funding from the EMBL. T.H. acknowledges the Hubrecht Institute for support, and the Hiiragi lab is also supported by the European Research Council (previously ERC Advanced Grant “SelforganisingEmbryo”, grant agreement 742732; currently ERC Advanced Grant “COORDINATION”, grant agreement 101055287), Stichting LSH-TKI (LSHM21020) and JSPS KAKENHI grant numbers JP21H05038 and JP22H05166. L.R.\ acknowledges the support of Italian National Group of Mathematical Physics (GNFM) of INDAM. L.R.\ also gratefully acknowledges support from the Italian Ministry of University and Research (MUR) through the grant PRIN2022-PNRR project (No. P2022Z7ZAJ) “A Unitary Mathematical Framework for Modelling Muscular Dystrophies” (CUP: E53D23018070001). The authors also express their gratitude to Fran\c{c}ois Graner for providing constructive feedback on the theoretical aspects of CPMs, as well as to Amitabha Nandi, Pamela Guruciaga, Jan Rombouts, Tim Dullweber, Jenna Elliott, Ergin Kohen, and Pietro Zamberlan for their feedback.
\end{acknowledgments}

\bibliography{main}

\pagebreak
\onecolumngrid
\vspace{3em}
\begin{center}\huge\textbf{Supplemental Materials}\end{center}
\vspace{-1em}

\renewcommand{\thesection}{\arabic{section}}
\renewcommand{\thefigure}{S\arabic{figure}}
\renewcommand{\theequation}{S\arabic{equation}}
\renewcommand{\thetable}{S\arabic{table}}

\titleformat{\section}
{\normalfont\Large\bfseries}{\thesection}{1em}{}
\titleformat{\subsection}
{\normalfont\large\bfseries}{\thesubsection}{1em}{}

\section{\label{sec:cpm}Poissonian dynamics of traditional CPMs}
By construction, CPMs describe spaces of discrete states, transitions between which represent stochastic jump processes. The modified Metropolis algorithm, traditionally adopted in the field, approximates the system's dynamics in a discretized time, and is manifestly Markovian—the probability of a given transition depends only on the current state of the system and has no memory of the past.

Given the Markovian nature of the CPM, we can invoke a classical result of queuing theory~\cite{Nelsen1987}: the distribution of waiting times between transitions must be exponential. The stochastic point process with exponential waiting times is Poissonian, which is therefore the most natural way to represent the dynamics of CPMs. To find such a representation we assume for simplicity that detailed balance is enforced in the implementation of the modified Metropolis algorithm as in Ref.~\cite{durandEfficientCellularPotts2016}.

To show how the modified Metropolis algorithm, regarded as a tau-leap approximation of a continuous Poisson process~\cite{Gillespie2001}, can be formulated using Eqs.~(17)--(18) we represent the state-transition probability of the $i$\textsuperscript{th} voxel $\sigma_i = s_0$ to  $\sigma_i = s_{m=1,2,...,M} \ne s_0$ within a time $\Delta{t}$ as $$
    p_{0m} = g_{0m} A_{0m} = 1 - e^{-k_{0m} \Delta{t}} \simeq k_{0m} \Delta{t},
$$
in which $g_{0m}$ and $A_{0m}$ are, respectively, the probability of proposal for a change $s_0 \to s_m$ and its acceptance probability, given the current state $\sigma_i = s_0$, and the $k_{0m}$ denote unknown Poissonian rates. Given that the traditional implementation proposes $P$ changes per one time step (drawn without replacement) and the voxel lattice of size $N$, we have
\begin{align}
    &g_{0m} = \frac{P}{M N},
    \\\label{eq:A0m}
    &A_{0m} = \min\{1, \exp[\beta\Delta{H}_{0m}]\}.
\end{align}
with $\beta = (k_B T)^{-1}$ and $\Delta{H}_{0m} = H(\sigma_i = s_0) - H(\sigma_i = s_m)$. Hence, the underlying Poissonian process can be reconstructed from
\begin{equation}\label{eq:k0m}
    k_{0m} = -\Delta{t}^{-1} \ln\left(1 - \frac{P}{M N}\min\{1, \exp[\beta\Delta{H}_{0m}]\}\right) \simeq \frac{P}{M N \Delta{t}}\min\{1, \exp[\beta\Delta{H}_{0m}].
\end{equation}

Equation~\eqref{eq:k0m} represents an implicit constraint imposed by the traditional CPMs through the choice of $g_{0m}$ and $A_{0m}$, which is \textit{not dictated by any physical laws or empirical observations}. Historically this choice was convenient to sample equilibrium ensembles when the dynamical details are ignored. However, there are infinitely many Poissonian processes, whose dynamics generates the same ensemble, and among which the traditional CPMs select implicitly only one.

Our framework affords a more general form of a Poissonian process given by Eqs.~(17) and (18). In equilibrium conditions, these processes generate a Boltzmann distribution with the designated Hamiltonian, whereas the action rates $\alpha_i(\bm{\sigma})$ can be arbitrary nonnegative functions of the state vector $\bm{\sigma}$. In the following, we assume that the Hamiltonian $H(\bm{\sigma})$ is given. With all voxels' states except $\sigma_i$ fixed, we denote for brevity $\alpha_0 = \alpha_i(\sigma_i = s_0 | \sigma_{j\ne i})$. Then the $i$\textsuperscript{th} voxel changes its current state with an escape probability [cf. Eqs.~(17)]
\begin{equation}\label{eq:escape}
    \kappa_0 = \sum_{m=1}^M k_{0m} =  \frac{\alpha_0}{\sum_{m=0}^M e^{\Delta{L}_{m0}}} \sum_{m=1}^M e^{\Delta{L}_{m0}}
        = \frac{\alpha_0}{1 + \sum_{m=1}^M w_m} \sum_{m=1}^M w_m,
\end{equation}
in which we introduced weights $w_m = \exp(\Delta{L}_{m0})$. The individual rates can then be expressed as
\begin{equation}\label{eq:k0m2}
    k_{0m} = \frac{\alpha_0 w_m}{1 + \sum_{n=1}^M w_n} = \frac{\kappa_0 w_m}{\sum_{n=1}^M w_n},
\end{equation}
which allows us to rewrite Eq.~(17) in the form
$$
    p(\sigma_i\to\sigma_i^{(m)}) = \frac{w_m}{\sum_{n=1}^{M} w_n} = \frac{k_{im}}{\kappa_0}.
$$
Furthermore, because the transition rate $k_{0m}$ and its reverse $k_{m0}$ satisfy the detailed-balance condition, we can derive a formula for the weights, which is analogous to Eq.~(18):
$$
    w_m = e^{\Delta L_{m0}} = \frac{\kappa_m}{\kappa_0} e^{-\beta \Delta{H}_{m0}},
$$
with $\kappa_m$ being the escape rate of the state $\sigma_i = s_m$. As follows from Eq.~\eqref{eq:k0m2} there exists a time constant
$$
    \tau_0 = \frac{\sum_m w_m}{\kappa_0} = \frac{\sum_m \kappa_m e^{-\beta \Delta{H}_{m0}}}{\kappa_0^2},
$$ such that $\tau_0 k_{0m} = w_m$. Hence from Eq.~\eqref{eq:escape} we can express the action rate in terms of $\tau_0$ as
$$
    \alpha_0 = \frac{1 + \kappa_0 \tau_0}{\tau_0}.
$$
In summary the above procedure requires as input the system's Hamiltonian and all the transition rates and yields the escape rates $\kappa_{j=0,1...M}$ and the time constants $\tau_j$, which then determine all action rates $\alpha_j$.

Whereas the dynamics of traditional CPMs implies a particular choice of the action rates $\alpha(\bm{\sigma})$, which can be reconstructed by following the above procedure, our approach offers their general functional form $\alpha_i(\bm{\sigma})$, which can be adjusted to relevant physical assumptions or learned from empirical data. In the main text we have proposed to set $\alpha_i(\bm{\sigma}) = \alpha_i(\sigma_i)$, which relies on local information about the current state of the $i$\textsuperscript{th} voxel. As discussed in the next section, this choice implies a constant response coefficient of cell surfaces to the thermodynamics forces.

Notably, the general theory of the Metropolis algorithm allows one to use any rule of the acceptance probability $A_{jm}$, as long as it ensures the detailed balance condition $p_{jm}/p_{mj} = (g_{jm} A_{jm}) / (g_{mj} A_{mj}) = \exp(\beta\Delta{H}_{jm})$. Using this property, one may replace the traditional rule \eqref{eq:A0m} by the softmax function
$$
    \tilde{A}_{jm} = \frac{e^{\beta \Delta{H}_{0m}}}{\sum_{n=1}^M e^{\beta \Delta{H}_{0n}}}.
$$
Then $g_{jm} = g_{mj}$ can be interpreted as the attempt probability with homogeneous action rates
$$
    \tilde{\alpha} = \Delta{t}^{-1} \ln(1 - g_{jm}) \simeq \frac{g_{jm}}{\Delta{t}} = \frac{P}{M N \Delta{t}}.
$$
Under these conditions the transitions $\tilde{p}_{jm} = g_{jm} \tilde{A}_{jm}$ satisfy the detailed balance condition [cf. Eqs.~(17) and (18)]. Hence, the Metropolis scheme can be modified to recapitulate special cases of the Poissonian dynamics, e.g. with homogeneous action rates $\tilde{\alpha}$.

\section{\label{sec:phi}Response coefficients and nonconservative forces}

The action rates $\alpha_i(\bm{\sigma})$ and active exponents $\phi(\sigma_i, \sigma_i^{(j)})$, introduced in the Poissonian framework of CPMs, can also be interpreted in the context of cell-surface mechanics~\cite{Mare,Magno2015,rensEnergyCellularForces2019}. In particular, action rates together with the spatial discretization give rise to the response coefficients of cell-surface elements, whereas the active exponents lead to nonconservative forces---not derived from the Hamiltonian---and amplify fluctuations, as shown below.

\begin{figure}
    \centering
    \includegraphics[width=0.4\textwidth]{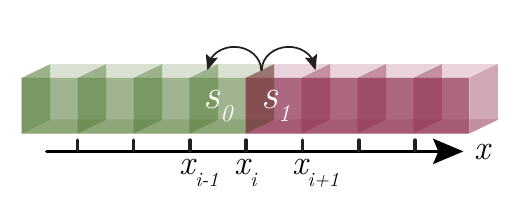}
    \caption{\label{fig:interface} Schematic of a surface element perpendicular to the $x$ axis at the interface between two states ($s_0$ and $s_1$), which undergoes a random walk.}
\end{figure}
Consider two voxels, $\sigma_0 = s_0$ and $\sigma_1 = s_1$, which share one face perpendicular to the spatial coordinate axis $x$, in different states $s_0 \ne s_1$ (Fig.~\ref{fig:interface}). Such a face represents an interface, e.g. between two cells, which is positioned with a probability $p(t, x_0) = 1$ at time $t$ and coordinate $x_0$. If at time $t + dt$ the voxel $\sigma_0$ ($\sigma_1$) copies the state from $\sigma_1$ ($\sigma_0$), the interface is moving left (right) on the discrete lattice to a new position $x_{-1} = x_0 - \Delta{x}$ ($x_{1} = x_0 + \Delta{x}$). For simplicity we assume that within some larger 3D neighborhood of $\sigma_0$ and $\sigma_1$ there are no voxels in other states $s \notin \{s_0, s_1\}$. Then we can express the probability of finding the interface at a given coordinate $x_i$ at time $t+dt$ as
\begin{equation}\label{eq:random}
    p(t + dt, x_i) = p(t, x_{i+1}) p_-(x_{i+1}) + p(t, x_{i - 1}) p_+(x_{i-1}) + p(t, x_i) [1 - p_+(x_i) - p_-(x_i)),
\end{equation}
in which $p_+(x_{i-1})$ and $p_-(x_{i+1})$ are the probabilities of the right and left moves, as determined by a change of the directing function $\Delta L_i = L(\sigma_i = s_0) - L(\sigma_i = s_1)$ in Eqs.~(17)--(18)
\begin{align}
    &p_+(x_{i-1}) \simeq dt \frac{\alpha(s_1) e^{\Delta{L}_i + \phi(s_1, s_0)}}{1 + e^{\Delta{L}_i + \phi(s_1, s_0)}}
    = dt \frac{\alpha(s_1) e^{\frac{\Delta{L}_i + \phi(s_1, s_0)}{2}}}{
        e^{-\frac{\Delta{L}_i + \phi(s_1, s_0)}{2}} + e^{\frac{\Delta{L}_i + \phi(s_1, s_0)}{2}}
    } = dt \frac{\sqrt{\alpha(s_0) \alpha(s_1)} e^{-\frac{\beta \Delta{H}_i - \phi(s_1, s_0)}{2}}}{
        e^{-\frac{\Delta{L}_i + \phi(s_1, s_0)}{2}} + e^{\frac{\Delta{L}_i + \phi(s_1, s_0)}{2}}
    },\\
    &p_-(x_{i+1}) \simeq  dt\frac{\alpha(s_0) e^{-\Delta{L}_i + \phi(s_0, s_1)}}{1 + e^{-\Delta{L}_i + \phi(s_0, s_1)}}
    = dt \frac{\alpha(s_0) e^{\frac{\phi(s_0, s_1) - \Delta{L}_i}{2}}}{
        e^{-\frac{\phi(s_0, s_1) - \Delta{L}_i}{2}} + e^{\frac{\phi(s_0, s_1) - \Delta{L}_i}{2}}
    } = dt \frac{\sqrt{\alpha(s_0)\alpha(s_1)} e^{\frac{\phi(s_0, s_1) + \beta\Delta{H}}{2}}}{
        e^{-\frac{\phi(s_0, s_1) - \Delta{L}_i}{2}} + e^{\frac{\phi(s_0, s_1) - \Delta{L}_i}{2}}
    }.
\end{align}
Given that the spatiotemporal scales $\Delta{x}$ and $dt$ are sufficiently small, the following expansions may be applied:
\begin{align}
    &e^{-\frac{\Delta{L}_i + \phi(s_1, s_0)}{2}} + e^{\frac{\Delta{L} + \phi(s_1, s_0)}{2}}
        \simeq 2 - \frac{\Delta{L}_i + \phi(s_1, s_0)}{2} + \frac{\Delta{L}_i + \phi(s_1, s_0)}{2} = 2,\\
    &e^{-\frac{\phi(s_0, s_1) - \Delta{L}_i}{2}} + e^{\frac{\phi(s_0, s_1) - \Delta{L}_i}{2}} \simeq 2.
\end{align}

Equation~\eqref{eq:random} defines a random walk, which has been thoroughly analyzed in the context of diffusion problems~\cite{Bringuier2011,Andreucci2018,Keller2004,Chandrasekhar1943,Belousov2016}. As we are following closely the notation of Ref.~\cite{Belousov2022}, we may immediately borrow its results to express
\begin{equation}\label{eq:master}
    \partial_t p = - \partial_x \left\{
        \frac{\Delta{x}}{dt} (p_+ - p_-) p - \frac{\Delta{x}^2}{2 dt} \partial_x\left[(p_+ + p_-)  p\right]
    \right\},
\end{equation}
into which we can substitute
\begin{align}
    p_+ - p_- \simeq& dt \frac{\sqrt{\alpha(s_0) \alpha(s_1)}}{2} \left[
        \frac{\phi(s_1, s_0) - \phi(s_0, s_1)}{2} - \beta\Delta{H}_i
    \right]
    = \frac{\beta dt \Delta{x}}{2} \sqrt{\alpha(s_0) \alpha(s_1)} \left(F - \frac{\Delta{H}}{\Delta{x}}\right),
    \\\label{eq:ppp}
    p_+ + p_- \simeq& dt\frac{\sqrt{\alpha(s_0) \alpha(s_1)}}{2} \left[2 + \frac{\phi(s_1, s_0) + \phi(s_0, s_1)}{2}\right]
        = \frac{dt}{2} \sqrt{\alpha(s_0) \alpha(s_1)} \left(2 + 2 \Phi\right) \simeq dt \sqrt{\alpha(s_0) \alpha(s_1)} e^{\Phi}.
\end{align}
Here we decomposed $\phi(s_1,s_0) = 2 \Phi + \beta F \Delta{x}$ and $\phi(s_0, s_1) = 2 \Phi - \beta F \Delta{x}$ into symmetric (frenetic) and asymmetric (entropic) parts, $\Phi$ and $F \Delta{x}$ respectivly, in which the latter represents the work done by the force $F$ when the interface moves right. In the continuous limit $\Delta{x}\to 0, dt\to0$, and Eqs.~\eqref{eq:master}--\eqref{eq:ppp} yield a Fokker-Planck equation~\cite{Belousov2022}
\begin{equation}\label{eq:diffusion}
    \partial_t p = - \partial_x \left[
        \beta D (F - \partial_x H) p + D e^\Phi \partial_x p
    \right],
\end{equation}
with the diffusion constant emerging as $\sqrt{\alpha(s_0)\alpha(s_1)}\Delta{x}^2 / 2 \to D$, and the gradient $\Delta{H}/\Delta{x} \to \partial_x H$. This diffusion equation stands in correspondence to a Langevin dynamics of the form~\cite{Lau2007}
\begin{equation}\label{eq:langevin}
    \dot{x} = \mu (F - \partial_x H) + \sqrt{2 D e^\Phi} dB,
\end{equation}
with the mobility coefficient $\mu = \beta D$ and the standard Brownian motion $dB$ [$\avg{dB} = 0$, $\avg{dB(t) dB(s)} = \delta(t - s)$].

A few remarks about Eqs.~\eqref{eq:diffusion} and \eqref{eq:langevin} are worth emphasizing:

\begin{itemize}
    \item The action rates together with the spatial discretization $\Delta{x}$ give rise to the response coefficient $\mu$ between thermodynamic forces and their conjugate currents.
    \item These equations contain a nonconservative force $F$, which does not stem from the Hamiltonian $H$ and emerges due to the antisymmetric part of the active exponents. This is a nontrivial extension of the CPMs afforded by the Poissonian framework. Our derivation also gives a statistical-physics justification of the conservative forces $-\partial_x H$ considered in Refs.~\cite{Mare,Magno2015,rensEnergyCellularForces2019}.
    \item The frenetic part of the active exponents amplifies noise and, thus, breaks the fluctuation-dissipation theorem. This effect can be summarised by the effective diffusion coefficient $D_\mathrm{eff} = D e^\Phi \ne k_B T \mu$.
    \item Equations~\eqref{eq:diffusion} and \eqref{eq:langevin} generalize the Smoluchowski equation and the Brownian dynamics in a potential $H(x)$~\cite{Belousov2022}, which emerge in the limit of vanishing nonequilibrium forces. Note, that in 1D the active force $F$ can always be effectively reincorporated into the Hamiltonian~\cite[Appendix~B]{Belousov2022}, but in more dimensions not all forces can be expressed through gradients of an effective Hamiltonian.
\end{itemize}

The relation between the diffusion coefficient $D \propto \alpha \Delta{x}^2$ and the spatial discretization $\Delta{x}$ has an important consequence. If one changes the spatial resolution $\Delta \tilde{x} = \xi \Delta{x}$, the action rates should also be rescaled to $\tilde{\alpha} = \alpha / \xi^2$ to preserve the transport properties of cell surfaces.

Lastly, we discuss the concept of effective temperature \cite{Cugliandolo1997,Polettini2019}, which one may invoke to account for the fluctuation-allowance parameter in the traditional CPMs. If we were to simulate the CPM at the effective temperature $T_\mathrm{eff} = k_B T e^\Phi$, with the detailed balance implied, the fluctuation-dissipation theorem would be restored at the expense of altering the response coefficient $\mu_\mathrm{eff} = D / (k_B T_\mathrm{eff}) = \mu e^{-\Phi}$, and thus either speed up or slow down the descent of the system along the energy gradients. This approach therefore misrepresents the kinetic details of the modeled systems.

However, by using an effective temperature, Poissonian CPMs are capable of extending the equilibrium fluctuation-dissipation theorem for the cell-surface dynamics consistently through manipulation of the transport coefficients. In particular, we set $T_\mathrm{eff} = k_B T e^\Phi$, and rescale all action rates $\alpha_\mathrm{eff} = \alpha e^{\Phi/2}$. Thereby the effective level of noise is amplified through $D_\mathrm{eff} = D(\alpha_\mathrm{eff}) = D e^\Phi$, whereas the mobility of cell-surface elements remains unaffected
$$\mu_\mathrm{eff} = D(\alpha_\mathrm{eff}) / k_B T_\mathrm{eff} = \mu.$$
Whereas the rescaling outlined above preserves the mesoscopic dynamics of cell surfaces by design, other mesoscopic and macroscopic variables may still be affected by these modifications~\cite{Polettini2019}.

\section{\label{sec:num}Computational details}
In all simulations of CPMs we used a 3D voxel grid with a resolution $\Delta{x}^3 = \SI{1}{\mu{m}^3}$ and a discrete tau-leap algorithm with a time step $\SI{0.1}{min}$ \cite{Gillespie2001}. To convert surface tension constants of the EPI and PrE cells with the medium, $\gamma_{\mathrm{medium}:\mathrm{EPI}}$ and $\gamma_{\mathrm{medium}:\mathrm{PrE}}$ respectively, to the interaction parameters $J_{\mathrm{medium}:\mathrm{EPI}}$ and $J_{\mathrm{medium}:\mathrm{PrE}}$ we used a formula $J_c = 6 \Delta{x}^{2} \gamma_c / 26$ for the contact type $c$, which can be interpreted as follows. A single voxel in our model has 6 faces of area $a = \Delta{x}^2$ and interacts with 26 sites in the surrounding Moore neighborhood. The surface energy of such a voxel in a medium is then $E_\mathrm{surface} = 6 a \gamma_c = 26 J_c$, from which the formula for $J_c$ follows. By using the same formula the tension constants of cell contacts $\gamma_{\mathrm{EPI}:\mathrm{EPI}}$, $\gamma_{\mathrm{EPI}:\mathrm{PrE}}$, $\gamma_{\mathrm{PrE}:\mathrm{PrE}}$ can also be converted to the spin interaction constants $J_{\mathrm{EPI}:\mathrm{EPI}}$, $J_{\mathrm{EPI}:\mathrm{PrE}}$, and $J_{\mathrm{PrE}:\mathrm{PrE}}$ respectively, and vice versa. The total surface area of a cell was energetically constrained~\cite{Graner2017}.

Because experimental observations at our disposal are not sufficient to fit parameters of our CPM, we had to estimate and adjust manually their values to achieve a similar dynamics of the sorting score in simulations. Thus we chose $J_{\mathrm{medium}:\mathrm{EPI}} = 30.0 \times 10^3 k_{\rm{B}} T$ and $J_{\mathrm{medium}:\mathrm{PrE}} = 9.2 \times 10^3 k_{\rm{B}} T$, which are close to the maximum experimentally observed tension $\gamma_{\mathrm{medium}:\mathrm{EPI}} = \SI{669}{pN / \mu{m}}$ and the minimum observed tension $\gamma_{\mathrm{medium}:\mathrm{PrE}} = \SI{178}{pN / \mu{m}}$ (see~\hyperref[sec:exp]{Experimental details}). Such a large difference of tension is necessary to compensate additional sorting mechanisms, which are not accounted for by our model and are discussed in the companion paper \cite{companion}. The cell-cell tension constants were chosen in the interval $(\gamma_{\mathrm{medium}:\mathrm{PrE}},\,\gamma_{\mathrm{medium}:\mathrm{EPI}})$ to favor the interface energy of PrE cells with the medium over EPI and homotypic contacts over heterotypic ones: $J_{\mathrm{EPI}:\mathrm{EPI}} = 18.5 \times 10^3 k_{\rm{B}} T$, $J_{\mathrm{EPI}:\mathrm{PrE}} = 25.4 \times 10^3 k_{\rm{B}} T$, and $J_{\mathrm{PrE}:\mathrm{PrE}} = 13.8 \times 10^3 k_{\rm{B}} T$. The experimental temperature is \SI{37}{\celsius}.

The volume elasticity constant $\kappa$ in Eq.~(1)
 is perhaps the most difficult parameter to choose, because we do not have experimental observations of cell-volume dynamics. A very large elasticity constant leads to artificial cuboidal cell shapes, whereas cells with a small elasticity and preferred volume may collapse to a single voxel due to the surface tension. Furthermore cells with a large value of $\kappa$ grow faster (see growth and division below). Combined with a large action rate such cells may acquire unrealistic shapes with long protrusions and no bulk volume. Empirically we found that $\kappa = 1154\, k_{\rm{B}} T / \Delta{x}^6$ allows us to observe realistic cell shape with action rates below $\SI{3.57}{min^{-1}}$.

At the present quite little is known about regulation of the cell mitosis and death in the inner cell mass of the mouse blastocyst. Experimentally we could however measure volumes of a few mitotic cells (see~\hyperref[sec:exp]{Experimental details}). Their mitotic cycle lasts approximately $\SI{12}{h}$ \cite{companion}. In the view of this information we decided to neglect the cell death and use the following model of growth and division.

\begin{figure}[ht]\centering
\includegraphics[width=0.9\textwidth]{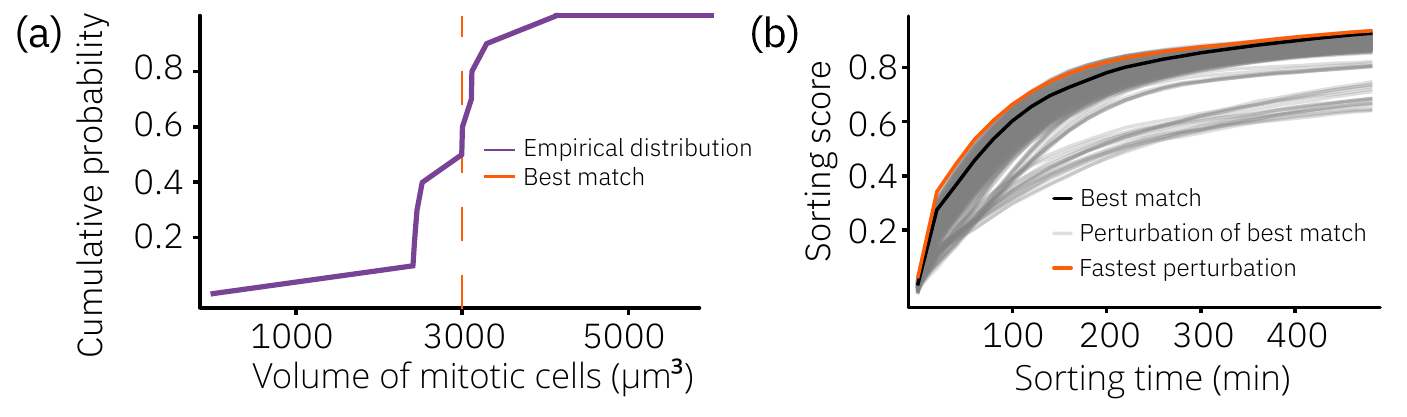}
\caption{\label{fig:si} Computational details: (a) Distribution of volume of mitotic cells in experiments (see \hyperref[sec:exp]{Experimental details}); (b) Modulation of sorting dynamics by action rates among 244 curves with all other parameters fixed as described in the text.}
\end{figure}

For each cell we sample a division volume $\tilde{V}$ from the empirical distribution found experimentally [Fig.~\ref{fig:si}(a)]. The cell's preferred volume $\tilde{V}$ is growing with a constant rate $g$ until the cell's \textit{actual} volume $V$ reaches or exceeds $\tilde{V}$. The mother cell is then divided approximately into two halves perpendicular to the major principal axis of the cell's gyration tensor \cite{KOPP2008,Kronenburg2013,Deledalle2017}.

From experiments we found that average volume of ICM cells is about $\SI{1500}{\mu{m}^3}$, whereas the average volume of mitotic cells is approximately $\SI{3000}{\mu{m}^3}$---the double of the average cell volume. Hence the cells should roughly double in size within one cell cycle. Given the chosen elasticity constant $\kappa$, we adjusted the growth rate of the preferred volume $g = \SI{3.36}{\mu{m}^3/min}$ to match the 12-hours duration of the mitotic cycle. The initial conditions for our simulations were generated from a single cell to match the observed median value of 25--26 cells per ICM at day E3.5, by setting the cell-medium and cell-cell tension constants and the cell action rate to the average values: $\gamma_{\mathrm{medium}:\mathrm{cell}} = (\gamma_{\mathrm{medium}:\mathrm{EPI}} + \gamma_{\mathrm{medium}:\mathrm{PrE}}) / 2$, $\gamma_\mathrm{cell:cell} = (\gamma_\mathrm{EPI:EPI} + \gamma_\mathrm{PrE:PrE} + \gamma_\mathrm{EPI:Pre}) / 2$, $\alpha_\mathrm{cell} = (\alpha_\mathrm{EPI} + \alpha_\mathrm{PrE})/2$. The cell fates were assigned randomly with 60\% PrE cells \cite{companion}.

Action rates were adjusted to match the sorting-score dynamics in experiments. Examples of dynamics modulated by the spanned range of action rates is illustrated in Fig.~\ref{fig:si}(b).

\begin{figure}[t]\centering
\includegraphics[width=0.8\textwidth]{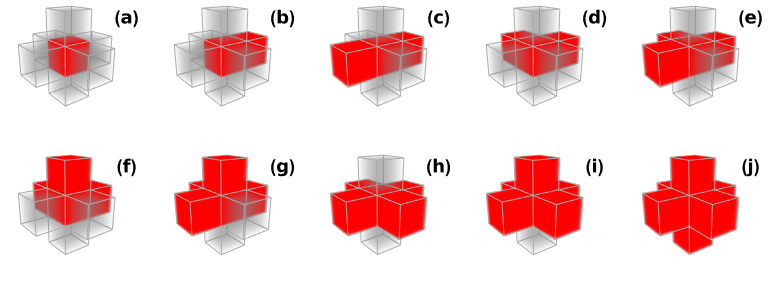}
\caption{\label{fig:adjacency} Scheme of the 3D adjacency neighborhood in the connectivity test of \textcite{durandEfficientCellularPotts2016} for 10 symmetrically nonequivalent configurations. Panel (a): the central site (red) contains no voxels of the same value in the six sites of the adjacency neighborhood (gray). If the red color represents a cell, the value of the central site cannot be changed, because the corresponding cell would disappear from the simulation. If the red color represents medium, in our simulations such a change may be admitted, although it would violate the detailed balance conditions and must be therefore interpreted as an active process, e.g. absorption of a fluid by cells. Panel (b): The central site has the same value as one of the six voxels in its adjacency neighborhood. Panels (c) and (d): The central site has the same value as two of the six voxels in its adjacency neighborhood. Panels (e) and (f): The central site has the same value as three of the six voxels in its adjacency neighborhood. Panels (g) and (h): The central site has the same value as four of the six voxels in its adjacency neighborhood. Whereas the case (g) requires the continuity test to ensure that changing the value of the central site is allowed, in the case (h) changes are not allowed, even though the red voxels in the adjacency neighborhood are connected. Panels (i) and (j): The central site has the same value as five and all six voxels of its adjacency neighborhood respectively. In the latter case no changes are allowed.}
\end{figure}

To prevent cell fragmentation we adopted the method of \textcite{durandEfficientCellularPotts2016} in a modified form. As we discussed with one of the two authors \cite{emailDurand}, their approach does not guarantee that the cells remain \textit{simply} connected in 3D models. In order to check that the cell is locally connected at a given site, we find all voxels occupied by the cell in the adjacency neighborhood of this site: all possible configurations of the adjacency neighborhood can be reduced to a total of 10 symmetrically nonequivalent situations to consider (Fig.~\ref{fig:adjacency}). One special case illustrated by Fig.~\ref{fig:adjacency}(h) passes the connectivity test described in Ref.~\cite{durandEfficientCellularPotts2016}, but changing the value of the central site in this configuration would create a hole. Therefore we prohibit such changes in our simulations. In the other situations we apply the method of Ref.~\cite{durandEfficientCellularPotts2016} in the original form.

The above modification of the connectivity test prevents a trivial scenario, in which one cell would ``drill'' a handle in another cell, whereas the connectivity of cells is always ensured by the test of \textcite{durandEfficientCellularPotts2016}. For the parameter values of interest we did not observe in our simulations appearance of holes or handles. Nonetheless our modification may not guarantee that such geometric defects would not occur for some extreme values of parameters. In 3D systems it appears that a hole at a given site could be created through a change of adjacent voxels.

\section{\label{sec:exp}Experimental details}

\subsection*{Cell tension measurements}

Micropipette aspiration was performed as described in \cite{maitre2015pulsatile} to measure the surface tension of ICM cells.
In brief, a micro-forged micropipette coupled to a microfluidic pump (Fluigent, Microfluidic Flow Control System) was used to measure the surface tension of ICM cells. Micropipettes were prepared from glass capillaries (Warner Instruments, GC100T-15) using a micropipette puller (Sutter Instrument, P-1000) and a microforge (Narishige, MF-900).
A fire-polished micropipette with a diameter of 7-8${\mu{m}}$ was mounted on an inverted Zeiss Observer Z1 microscope with a CSU-X1M 5000 spinning disc unit, and its movement was controlled by micromanipulators (Narishige, MON202-D). Samples were maintained at \SI{37}{\celsius} with 5\% $CO_2$. A stepwise increasing pressure was applied on ICM surface cells using the microfluidic pump and Dikeria software (LabVIEW), until a deformation with the same radius as that of the micropipette ($R_p$) was reached. The equilibrium pressure ($P_c$) was measured, images were acquired in this configuration and then the pressure was released. At steady state, the surface tension $\gamma$ at the cell-fluid interface is calculated based on Young–Laplace’s law: $\gamma = P_c/2(1/R_p - 1/R_c)$, in which $P_c$ is the net pressure used to deform the cell of radius $R_c$. Image analysis and measurement of the pipette radius $R_p$ and $R_c$ was done in FIJI, and calculation of surface tension was done using custom scripts in Python v3.9.

\subsection*{Live-imaging of externalized inner cell mass}

For externalization of the inner cell mass from mouse blastocysts, the zona pellucida (ZP) was first removed from blastocysts with pronase (0.5\% w/v Proteinase K, Sigma P8811, in global medium containing HEPES (LifeGlobal, LGGH-050 supplemented with 0.5\% PVP-40, Sigma, P0930) treatment for 2-3 minutes at \SI{37}{\celsius}. Blastocysts were washed in 10$\mu$l droplets of global medium (LifeGlobal, LGGG-050). To isolate the inner cell mass, blastocysts were incubated in serum containing anti-mouse antibody (Cedarlane, CL2301, Lot no. 049M4847V) diluted 1:3 with global medium for 30 minutes at \SI{37}{\celsius}. Following 2-3 brief washes in global medium with HEPES, embryos were incubated in guinea pig complement (Sigma, 1639, Lot no. SLBX9353) diluted with global medium in a 1:3 ratio for 30 minutes at \SI{37}{\celsius}. Lysed outer cells and remaining debris were removed by gentle pipetting with a narrow glass capillary (Brand, 708744) to isolate the inner cell mass. The isolated ICMs were cultured in 10$\mu$l drops of global medium in a petri dish (Falcon, 351008) covered with mineral oil (Sigma, M8410) and incubated at \SI{37}{\celsius} with 5\% $CO_2$ for up to 24 hours.

Externalized ICMs were placed into global medium (LifeGlobal, LGGG-050) drops covered with mineral oil on a glass-bottom imaging dish (MatTek, P50G-1.5-14-F). Time-lapse imaging of live, fluorescent samples was performed on an inverted Zeiss Observer Z1 microscope with a CSU-X1M 5000 spinning disc unit. Excitation was achieved using 488 nm, and 561 nm laser lines through a 63/1.2 C Apo W DIC III water immersion objective. Emission was collected through 525/50nm, 605/40nm, band pass filters onto an EMCCD Evolve 512 camera. Images were acquired every 20 minutes for up to 12 hours. The microscope was equipped with a humidified incubation chamber to keep the samples at \SI{37}{\celsius} and supply the atmosphere with 5\% $CO_2$.

\subsection*{Immunostaining and imaging of ICMs}
Externalized ICMs were fixed in 4\% PFA (Sigma, P6148) at room temperature
for 15 minutes. Fixed ICMs were washed 3 times (5 minutes each) in wash buffer
(PBS-Tween containing 1\% BSA), and permeabilized at room temperature for 20
minutes in permeabilization buffer (0.5\% Triton-X in PBS; Sigma T8787). After permeabilization, samples were washed (3 x 5 minutes), followed by incubation in blocking buffer (PBS-Tween containing 3\% BSA) for 2 hours at room temperature. Blocked samples were then incubated with desired primary antibodies diluted 1:200 in blacking buffer overnight at 4°C. Next, the samples were washed (3 x 5 minutes), and incubated in fluorophore-conjugated secondary antibodies and dyes diluted 1:200  in wash buffer at room temperature for 2 hours. Stained samples were washed (3 x 5 minutes) and incubated in DAPI solution (Life Technologies, D3571; diluted 1:1000 in PBS) for 10 minutes at room temperature. These samples were then transferred into individual droplets of PBS covered with mineral oil on a 35mm glass bottom dish (MatTek, P35G-1.5-20-C) for imaging. Primary antibodies against GATA6 (R\&D Systems, AF1700), GATA4 (R\&D Systems, BAF2606), NANOG (ReproCell, RCAB002P-F), and SOX2 (Cell Signaling Technology, 23064) were used. Secondary antibodies donkey anti-goat Alexa Fluor 488 (Thermofisher, A11055) and donkey anti-rabbit Alexa Fluor 647 (ThermoFisher, A31573) were used in this study. Phalloidin Rhodamine (Invitrogen, R415) was used at 1:200 to visualize actin.
 
Immunostained samples were imaged on a Zeiss LSM880 microscope with Airy-
Scan Fast mode. A 40× water-immersion Zeiss C-Apochromat 1.2 NA objective was
used, and raw Airyscan images were acquired and processed using the ZEN black
software (Zeiss).

\subsection*{Volume distribution of cells}

Cell volume measurement was performed using immunofluorescence images of externalized ICMs at different stages of development in Imaris v9.7.2 (Bitplane). The Cell module in Imaris was used to semi-automatically segment individual cells based on the fluorescence signal of their membranes. This was followed by a manual curation step to validate the segmentation. Correctly segmented cells were chosen for volume measurements.

\subsection*{Aspect-ratio distribution of cells}

Cell aspect ratio was measured from 3D time-lapse images of the ICM in whole blastocysts. The blastocysts were chimeras that were sparsely labeled for membrane fluorescence (mTmG) and a cell-fate reporter (\textit{Pdgfra\textsuperscript{H2B-GFP}}) in a few cells to visualize cell shape. The fluorescently labeled cells were tracked in time, and cell shape was manually traced in the equatorial maximum-intensity plane of the cell at each timepoint using FIJI. The cells were classified as PrE or EPI from the corresponding nuclear fluorescence signal of the fate marker \textit{Pdgfra\textsuperscript{H2B-GFP}}. From the resulting shape descriptors we calculated cellular aspect ratios of the shorter axis to the longer one.

\end{document}